\documentclass[a4paper,fleqn,usenatbib]{mnras}
\usepackage{newtxtext,newtxmath}

\usepackage[T1]{fontenc}
\usepackage{ae,aecompl}

\usepackage{graphicx}	
\usepackage{amsmath}	
\usepackage{amssymb}	

\title[Case AD, AR, and AS binary evolution]{Case AD, AR, and AS binary evolution and their possible connections with W UMa binaries}

\author[D. Jiang]{
Dengkai Jiang,$^{1,2,3}$\thanks{E-mail: dengkai@ynao.ac.cn}
\\
$^{1}$Yunnan Observatories, Chinese Academy of Sciences, 396 Yangfangwang, Guandu District, Kunming, 650216, P.R. China\\
$^{2}$Center for Astronomical Mega-Science, Chinese Academy of Sciences, 20A Datun Road, Chaoyang District, Beijing, 100012, P.R. China\\
$^{3}$Key Laboratory for the Structure and Evolution of Celestial Objects, Chinese Academy of Sciences, Kunming, 650011, China
}

\date{Accepted XXX. Received YYY; in original form ZZZ}

\pubyear{2015}

\begin{document}
\label{firstpage}
\pagerange{\pageref{firstpage}--\pageref{lastpage}}
\maketitle

\begin{abstract}
Close detached binaries were theoretically predicted to evolve into contact by three subtypes of case A binary evolution: case AD, AR, and AS, which correspond to the formation of contact during dynamic-, thermal-, and nuclear-timescale mass transfer phase, respectively. It is unclear, however, what is the difference between contact binaries in these subtypes, and whether all of these subtypes can account for the formation of observed W UMa binaries. Using Eggleton's stellar evolution code with the nonconservative assumption, I obtained the low-mass contact binaries produced by case AD, AR, and AS at the moment of contact, and their parameter spaces. The results support that the progenitors of low-mass contact binaries are detached binaries with orbital periods shorter than $\sim2-5\,$d, and their borderlines depend strongly on the primary mass. In addition, the period-colour relations for case AR and AS can be in better agreement with that for observed W UMa candidates, but case AD shows a significantly worse agreement. Moreover, case AR and AS can produce a short-period limit (corresponding to a low-mass limit) at almost any age, e.g. from young age ($\sim0.2\,$Gyr) to old age ($\sim13\,$Gyr), agreeing with observed W UMa binaries in star clusters, but no such limit occurs for case AD at any age. These results support that case AR and AS, as opposed to case AD, can lead to W UMa binaries (including young W UMa binaries). 


\end{abstract}




\section{Introduction}

Low-mass contact binaries have a peanut shape where two stars fill their respective Roche lobes. Low-mass contact binaries are important in understanding the formation and evolution of close binaries, because they contain the least amounts of angular momentum among all binaries without compact components (black holes, neutron stars, or white dwarfs). Mass and energy can be transferred between two stars in low-mass contact binaries \citep{Li 2004a, Li 2004b, Yakut 2005, Jiang 2009, Kouzuma 2018}, and therefore they attract special attention as laboratories to study these physical processes in close binaries. Low-mass contact binaries are the progenitors of the only stellar merger  (V1309 Sco) observed before \citep{Tylenda 2011}, which make them an important source to understand binary merger \citep{Webbink 1976,Hut 1980, Jiang 2010, Stepien 2011}. As an important part of blue straggler population \citep{Rucinski 1998, Rucinski 2000}, low-mass contact binaries contain the information on the dominant formation mechanism of blue stragglers, because low-mass contact binaries were observed in both blue-straggler sequences in globular clusters \citep{Ferraro 2009, Dalessandro 2013}, which suggests that binary evolution can contribute to both blue-straggler sequences \citep{Jiang 2017}. Moreover, the  absolute magnitudes of low-mass contact binaries are related to their orbital periods and colours, implying that they can be used as a distance tracer to investigate Galaxy structure \citep{Rucinski 1994, Chen 2016, Mateo 2017, Chen 2018}.

The most well-known observable counterpart of low-mass contact binary is W Ursae Majoris (W UMa) type eclipsing binary. Their defining feature is the sinusoidal-like light curve with equal-depth minima. This suggests that two components have similar temperatures, although they may have very different masses revealed by spectroscopic observation. W UMa binaries are the most numerous eclipsing binaries, and have been detected in any properly observed stellar system, for example, the field, open clusters \citep[e.g.][]{Rucinski 1998, Sriram 2009, Luo 2015, Chen 2016aj}, globular clusters \citep[e.g.][]{Rucinski 2000, Likai 2017}. Tens of thousands of candidates have been discovered from surveys, e.g. All Sky Automated Survey \citep{Pojmanski 2005}, the Optical Gravitational Lensing Experiment \citep{Rucinski 1996}, Super Wide Angle Search for Planets \citep{Norton 2011}, \textit{Kepler} space mission \citep{Prsa 2011, Slawson 2011} , and Catalina Real-Time Transient Survey \citep{Drake 2014a, Drake 2014b, Drake 2017}. Their spatial frequency relative to FGK type main-sequence stars is $\sim0.2\%$ in the solar neighborhood established from the \emph{Hipparcos} and ASAS samples \citep{Rucinski 2002, Rucinski 2006}, while their frequency was found to be very high in open clusters \citep{Rucinski 1994} and globular clusters \citep{Rucinski 2000}. Recently, \citet{Chen 2018} had compiled a catalog of 55,603 contact binary candidates, and found that their luminosity function may be different in different environments of the Galaxy.

The origin of W UMa binaries is one of the most fundamental questions, because detailed knowledge of their origin is required by many unsolved problems in the investigation of W UMa binaries, such as their over-luminosity of the secondaries \citep{Lucy 1968}, their short-period limit \citep{Rucinski 1992, Stepien 2006, Jiang 2012}, and their structure and evolution \citep{Kahler 2002a, Kahler 2002b, Kahler 2004, Li 2004a, Li 2004b, Li 2005}. At present, the detached-binary channel is one of the popular formation channels. Various studies and discussions have been emerged since 1960s, e.g. \citet{Huang 1966, Roxburgh 1966, Webbink 1976, van't Veer 1979, Vilhu 1982, Stepien 1995, Webbink 2003, Tutukov 2004, Jiang 2014MN}. In this channel, a close binary may evolve into semi-detached phase by evolutionary expansion of the components \citep{Webbink 1977a} and angular momentum loss \citep{Vilhu 1982}, and subsequently evolves into contact when the accretor star fills its Roche lobe in response to mass transfer. Another main formation mechanism of W UMa binaries is fission process, which can produce a W UMa binary directly at the end of the pre-main-sequence contraction (Roxburgh 1966), and therefore can explain young (<0.5\,Gyr) W UMa binaries \citep{Bilir 2005}. However, it is still unclear which is the dominant formation mechanism of young W UMa binaries, because the detached-binary channel can also produce such young W UMa binaries \citep{Jiang 2014MN}.

In the detached-binary channel, the progenitors of W UMa binaries (sometimes referred to as "precontact binaries") are usually detached or semidetached binaries with EA- or EB-type light curves (unequal eclipse depths). They may be chromospherically very active and may exhibit even the flares. Many close binaries have been observed with the determination of fundamental stellar parameters \citep[e.g.][]{Yakut 2005, Matson 2017}, and stellar activities have also been detected in many close binaries \citep[e.g.][]{Balona 2015, Gao 2016, Zhang 2018, Luo 2019}. Some close binaries were observed as progenitors of W UMa binaries \citep[e.g.][]{Kaluzny 1990, Maceroni 1999, Samec 2013}. At present, many observational studies mainly focused on detached binaries with periods shorter than 1d as the progenitors of W UMa binaries. However, theoretical predictions suggested that the progenitors of W UMa binaries may have an upper-period limit much longer than 1\,d, e.g. $\sim2-4\,$d \citep{Vilhu 1982, Stepien 2011, Jiang 2014MN}. In addition, the mass ratio may be another important parameter to determine whether or not a detached binary is the progenitors of W UMa binaries, but the mass-ratio range of the progenitors still remains unclear.


 It should be noted that even after low-mass close binaries evolve into contact, there is a long-standing challenge of how these contact binaries evolve. Lucy (1976) and Flannery (1976)  predicted that contact binaries undergo thermal relaxation oscillations (TRO) about a state of marginal contact \citep[see also][]{Webbink 1976, Webbink 1977a, Webbink 1977b, Sarna 1989, Yakut 2005}. Each oscillation includes a contact phase and a semi-detached (non-contact) phase, and this periodic evolution process is driven by mass and energy transfer between two components. In the contact phase, energy transfer is from the massive component to the less massive component, which leads to mass transfer in the opposite direction. Mass transfer from the less massive component to the more massive component will drive the evolution from contact to semi-detached phase. In the semi-detached phase, the more massive components transfer mass to the less massive component, driving the system back into a contact phase. This model can successfully explain observed properties of W UMa binaries, e.g. the observed light curves.

Previous studies \citep{Eggleton 2000, Nelson 2001, Eggleton 2006} indicated that three evolutionary routes of case A binary evolution can bring low-mass detached binaries into contact with two main-sequence components. They evolve into contact during the dynamical-timescale mass transfer phase (dynamic RLOF, case AD), during the thermal-timescale mass transfer phase (rapid to contact, case AR), or during the nuclear-timescale mass transfer phase (slow to contact, case AS), respectively. \citet{Nelson 2001} presented the initial parameter spaces of these subtypes with conservative assumption of total mass and orbital angular momentum, but they suggested that the nonconservative assumption is needed to be investigated because the conservative theoretical models were a worse agreement with the observed cool Algols. Moreover, different evolutionary routes of pre-contact phase may probably produce a contact binary with different nature. For example, \citet{Nelson 2001} suggested that case AR may produce a W UMa type binaries, while case AD is likely to lead to a binary merger after a short-timescale phase (contact or common envelope phase). Based on this suggestion, \citet{Jiang 2012} proposed dynamic mass transfer instability to explain the short-period limit of W UMa binaries. However, additional evidence is needed to support this suggestion about the relations between W UMa binaries and three subtypes of case AD, AR, and AS. Many factors are also still poorly known for three subtypes with the nonconservative assumption, such as their initial parameter space and their timescales of pre-contact phase. Therefore, more detailed work with the nonconservative assumption also needs to fully investigate these three subtypes.

The purpose of this paper is to investigate the formation of low-mass contact binaries from detached binaries through case AD, AS, and AR with the nonconservative assumption, and to investigate their possible connections with W UMa binaries. The outline of this paper is as follows. The binary evolutionary calculations are described in Sect. 2. In Sect. 3, the binary evolutionary results are presented.  The discussion and conclusions are given in Sect. 4 and Sect. 5.

\section{Binary evolution calculations}

To determine whether a detached binary can evolve into contact when two components are still MS stars, it is necessary to perform detailed binary evolution calculations. Eggleton's stellar evolution code (Programme EV, private communication 2003)\footnote{This version of the code with a user manual can be obtained on request from ppe@igpp.ucllnl.org or Peter.Eggleton@yahoo.com.} is employed in this paper. This code was originally developed by \citet{Eggleton 1971, Eggleton 1972, Eggleton 1973}, which have been updated over the last four decades \citep{Han 1994, Pols 1995, Nelson 2001, Eggleton 2002, Yakut 2005, Eldridge 2008}. This version of code includes the models of the nonconservative effects, e.g. dynamo-driven mass loss, magnetic braking, and tidal friction to the evolution of stars with cool convective envelopes \citep{Eggleton 2002, Yakut 2005, Eggleton 2006, Eggleton 2010}.

In this version of the EV code, there are two different modes for calculating the evolution of binaries: normal (non-TWIN) mode and TWIN mode. In the normal mode, the stellar equations for the mass donor (the primary) are first solved, and its mass-transfer history is stored. Then, this history is used as input for a subsequent calculation of the evolution of the mass gainer (the secondary). This mode has been used to compute Case A binary evolution with conservative assumption \citep{Nelson 2001}. In the TWIN mode developed by P. P. Eggleton in April 2002, the stellar equations of both stars in a binary are solved simultaneously. Therefore, this mode can be used to model heat transfer between two stars \citep{Yakut 2005} and non-conservative mass transfer in massive close binaries \citep{de Mink 2007}. To study three subtypes of case AD, AR, and AS with the nonconservative assumption, the TWIN mode is used in this work to solve both stars simultaneously. This is because the orbit of the primary will be varying due to the nonconservative effects on the secondary, if both stars are subject to the nonconservative effects of dynamo-driven mass loss, magnetic braking, and tidal friction \citep{Eggleton 2002, Yakut 2005, Eggleton 2006, Eggleton 2010}.

The calculation of mass-transfer rate between two stars in the TWIN mode is also different from that in the normal mode. In the normal mode, the mass-transfer rate for one star filling its Roche lobe is only treated as a function of its stellar radius and its Roche lobe radius \citep{Nelson 2001}. However, an alternative way of computing mass-transfer rate is needed when both stars are solved simultaneously. Therefore, mass-transfer rate between two stars in the TWIN mode is treated as a function of the surface potentials and the Roche surface potentials for both stars. The direction of mass transfer depends on the potential difference between stellar surfaces of two stars, and the mass-transfer rate is computed by the integral of a function of the potential difference between the surface of each layer and the Roche-lobe surface for all layers outside the Roche lobe  \citep[more details see][]{de Mink 2007}.

The evolutionary tracks were calculated with different initial primary masses $M_{10}$, mass ratio $q_0= M_{10}/M_{20}>1$, and orbital periods $P_0$ at the metallicity of Z = 0.02.These initial parameters are at equal logarithmic intervals, 
\begin{equation}
\log M_{10} = -0.35, -0.3,..., 0.6,
\end{equation}
\begin{equation}
\log (q_{0}) = 0.05, 0.10,..., 1.0,
\end{equation}
\begin{equation}
\log (P_0/P_\text{ZAMS}) = 0.05, 0.10,...,
\end{equation}
where $P_\text{ZAMS}$ is an approximation of the orbital period where the primary fills its Roche lobe on zero-age main sequence for a system with equal masses \citep{Nelson 2001}.   The range of the primary masses is chosen from 0.45 to 4\,M$_\text{\sun}$ to focus on the formation of low-mass contact binaries. Many low-mass binaries with primary mass $<$0.9\,M$_\text{\sun}$ may reach RLOF with the nonconservative assumption, although they can not reach RLOF in 20\,Gyr with conservative assumption \citep{Nelson 2001}. The upper limits of the initial orbital period are chosen to cover case AD, AR, and AS binary evolution for different primary mass.

In the calculations of these binaries, both stars in a binary start to evolve simultaneously from zero-age main sequence. When the main-sequence primary fills its Roche lobe (classified as case A), its matter will be transferred to the secondary. If the mass transfer of a binary accelerates to the dynamic timescale rapidly (e.g. $\dot M >M/t_\text{dyn}$), this binary would be expected to experience dynamic RLOF and is classified as case AD in this work, following the classification of \citet{Eggleton 2000} and \citet{Nelson 2001}. A binary is classified as case AR (rapid evolution to contact) when the secondary fills rapidly its Roche lobe as a result of mass accretion during thermal-timescale ($t_\text{KH}$) mass transfer phase (AR, $M/t_\text{dyn}>\dot M >M/t_\text{KH}$). If a binary evolves into contact not during the thermal-timescale mass transfer phase, but during the following nuclear-timescale mass transfer phase ($\dot M <M/t_\text{KH}$), it is classified as case AS (slow evolution to contact). 

In this paper, the mass-transfer rates and thermal timescales of the binaries are determined in the calculation of binary evolution, and their dynamic timescales are assumed to be less than one-tenth of the thermal timescale given by \citet{Nelson 2001}. To compare with the results given by \citet{Nelson 2001}, these binaries are calculated to evolve to 20\,Gyr unless the code experiences convergence problems, thus making the evolutionary calculations stop. The evolutionary calculation of a binary also stops at the moment when it is classified as case AD, AR, or AS, because this paper only focused on the formation of contact binaries, not their evolution.



\section{Binary evolution results}


\subsection{The initial parameter spaces of case AD, AR, and AS}

\begin{figure*}
	\includegraphics[width=\textwidth]{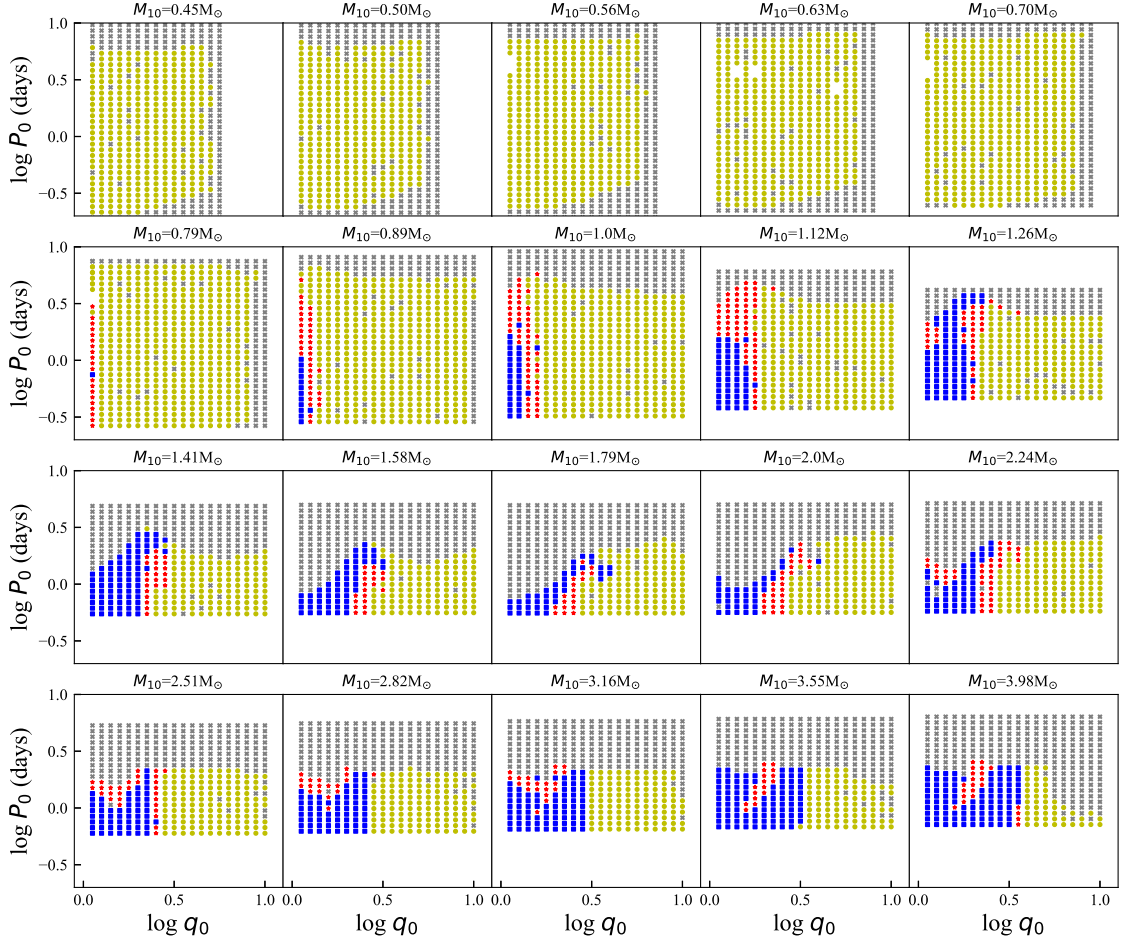}
    \caption{Initial parameter spaces of low-mass contact binaries for case AD (yellow filled circles), AR (red filled stars), and AS (blue filled squares) in the log$P_\text{0}-$log$q_\text{0}$ plane for different values of $M_\text{10}$.  Grey crosses denote systems that can not evolve into contact while both components are main-sequence stars.}
    \label{fig1}
\end{figure*}

The initial parameter space is important for investigating case AD, AR, and AS, but is still poorly known. In Fig.~\ref{fig1}, the initial parameters were mapped in the orbital period-mass ratio plane (for a range of primary masses), which lead to a contact binary through case AD, AR, and AS. It can be seen that contact binaries from these three subtypes have different initial parameter spaces and the initial primary mass has a strong influence on the initial parameter spaces of case AD, AR, and AS. With the decrease of primary mass, the parameter space of case AD becomes larger. In contrast, the parameter spaces of case AR and AS become smaller, and disappear finally, although the upper limits of the period for case AR and AS increases from $P_0=2$\,d at 3.98\,M$_\text{\sun}$ to $P_0=5$\,d at 1.0\,M$_\text{\sun}$. For the same primary mass, case AD occurs in the binaries with larger $q_0$, while case AR and AS occur in the binaries with smaller $q_0$. This initial parameters space can be used to estimate the outcome of a detached binary, or to find the progenitors of contact binaries for case AD, AR, and AS. If the parameters of a detached binary are located in these regions, a low-mass contact binary is then expected to be produced by case AD, AR, or AS in future.

\subsection{The possible connections between three subtypes and W UMa binaries}



\subsubsection{The period-effective temperature relations for case AD, AR, and AS}

\begin{figure*}
	\includegraphics[width=\textwidth]{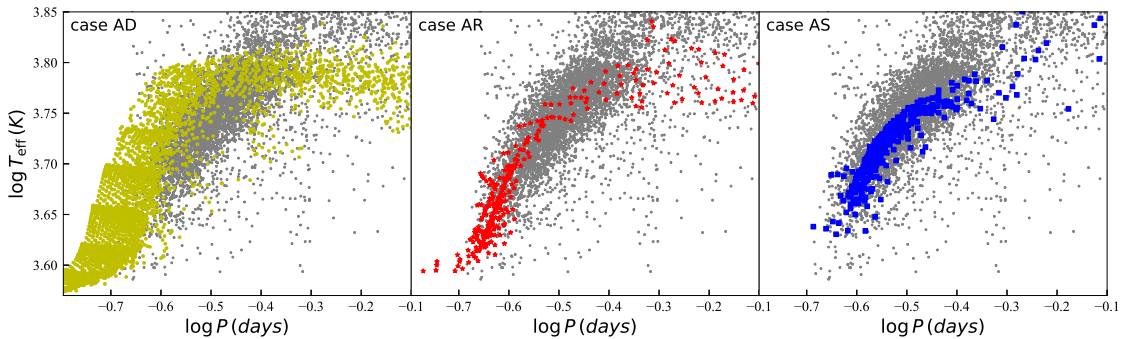}
    \caption{The distribution of low-mass contact binaries in the log$P-$log$T_\text{eff}$ diagram. Yellow filled circles, red filled stars, and blue filled squares represent case AD, AR, and AS, respectively. Gray points are observed W UMa binaries candidates compiled by \citet{Chen 2018}.}
    \label{fig2}
\end{figure*}

The period-colour relation (sometimes used the period-effective temperature relation) as found by \citet{Eggen 1967} has played a critical role in the testing of models of W UMa systems ever since \citep{Li 2004a, Li 2004b}. Based on a recent catalog of W UMa binaries candidates compiled by \citet{Chen 2018}, the possible connections were investigated between W UMa binaries and three subtypes (case AD, AR, and AS). Figure~\ref{fig2} shows these observed candidates of W UMa binaries and the low-mass contact binaries from case AD, AR, and AS in the plane of log($P-T_\text{eff}$). The effective temperatures of these observed candidates were taken by cross-matching with the A, F, G, and K type star catalog from the LAMOST DR5 data \citep{Cui 2012, Deng 2012, Luo 2012, Zhao 2012} using a 2\arcsec angular radius. Traditional contact binary models assumed energy transfer between two stars to explain their nearly equal temperatures. Following \citet{Jiang 2014MN}, the effective temperatures of low-mass contact binaries produced by case AD, AR, and AS can be corrected for this energy transfer according to $T^4=(R_1^2T_1^4+R_2^2T_2^4)/(R_1^2+R_2^2)$, where $R_1, R_2, T_1$, and $T_2$ are the radii and the effective temperatures of the primary and the secondary at the moment that the binary systems evolve into contact. 

As shown in Fig.~\ref{fig2}, the low-mass contact binaries with longer orbital period have higher temperature in case AD, AR, and AS, which is in agreement with the observed trend of W UMa binaries candidates. However, there is a disagreement between case AD and the observed W UMa candidates. Case AD can not cover the observed W UMa candidates with log$P<-0.5$, and case AD has much higher temperature than observed systems with the same orbital period. Case AD also shows a clearly flat distribution for log$P>-0.5$, which is also not consistent with observed distribution. Case AR and AS are in much better agreement with the observed sample than case AD. They cover the observed W UMa candidates with log$P<-0.5$, and are much closer to the observed sample for log$P>-0.5$. Therefore, the period-effective temperature (or colour) relations for case AR and AS can be in better agreement with that for observed W UMa candidates, but case AD shows a significantly worse agreement.

\subsubsection{The short-period limits and the pre-contact timescales for case AD, AR, and AS}

\begin{figure*}
	\includegraphics[width=\textwidth]{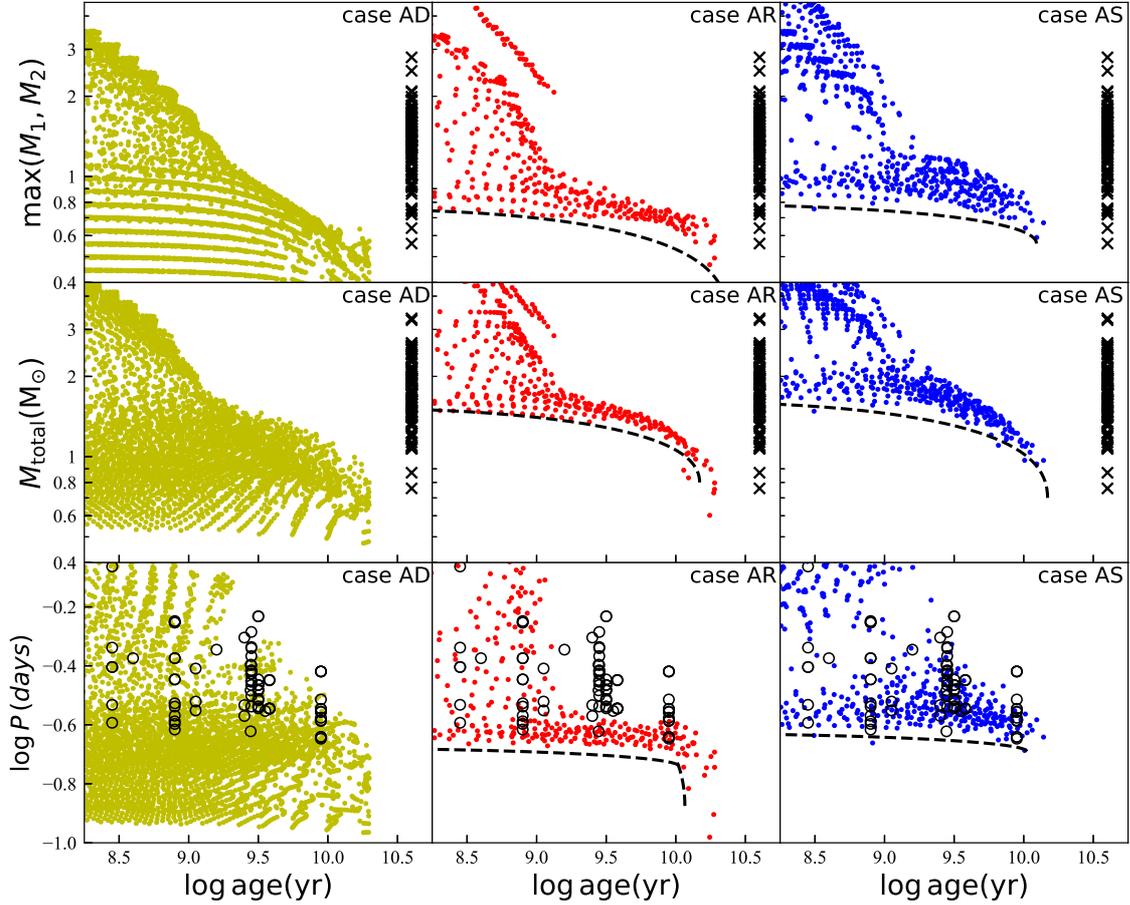}
    \caption{The distribution of the masses of the more massive components (upper panels), total masses (middle panels) and orbital period (lower panels) of low-mass contact binaries when they evolve into contact through case AD, AR, and AS. Black crosses in the upper and middle panels show the observed W UMa sample compiled by \citet{Yildiz 2013}, and open circles in the lower panels are observed W UMa sample in star clusters compiled by \citet{Bukowiecki 2012}.}
    \label{fig3}
\end{figure*}

The short-period limit (also corresponding to a low-mass limit) is one of the most important characteristics of W UMa binaries \citep{Rucinski 1992}, which can further constrain the possible connections between three subtypes and W UMa binaries. Figure~\ref{fig3} shows the mass of the more massive component, the total mass, and the orbital period of low-mass contact binaries when they evolve into contact through case AD, AR, and AS, and the age presents the pre-contact timescales (the timescales from a detached-binary formation to contact). As shown in this figure, all three subtypes can produce low-mass contact binaries in a wide range of age from 0.1\,Gyr to 15\,Gyr , and this agrees with the previous suggestion that the formation of contact binaries from detached binaries occur at any age \citep{Li 2007, Jiang 2014MN}. However, their distributions of the mass of the more massive component and the total mass are significantly different as shown in the upper and middle panels of Fig.~\ref{fig3}. Case AR and AS show the existence of a low-mass limit (the dashed lines in Fig.~\ref{fig3}), while case AD does not show such a low-mass limit. These low-mass limits exist in case AR and AS at almost any age, although they decrease with age. These low-mass limits were compared with the observed W UMa sample with well-determined parameters compiled by \citet{Yildiz 2013} from \citet{Yakut 2005, Gazeas 2005, Gazeas 2006, Zola 2005, Zola 2010}. The minimum values of the observed W UMa sample are 0.56\,M$_\text{\sun}$ and 0.76\,M$_\text{\sun}$ for the primary mass and total mass, respectively. These values seem to agree with the low-mass limits for case AR and AS, although there is no information about the age of these observed W UMa samples.

It should be noted that star clusters can offer information on the age of W UMa sample observed in star clusters, because the age of observed W UMa sample is the same as the cluster age. \citet{Bukowiecki 2012} collected W UMa binaries observed in star clusters and listed their orbital periods and ages, and their sample can be used to investigate the possible connections between three subtypes and W UMa binaries. These observed samples were compared with low-mass contact binaries produced by case AD, AR, and AS in lower panels of Fig.~\ref{fig3}. It can be seen that case AR and AS predict that low-mass contact binaries should have a short-period limit at any age, similar to a low-mass limit as shown in the upper and middle panels of Fig.~\ref{fig3}, while there is no such a short-period limit for case AD at any age. The most interesting result is that observed W UMa samples in clusters also show a short-period limit at a different age, agreeing with that predicted by case AR and AS. The existence of a short-period limit of W UMa binaries at a different age provides another strong evidence that case AR and AS, as opposed to case AD, can lead to W UMa binaries (including very young W UMa binaries).

\section{Discussion}

\subsection{Comparison with the models of \citet{Nelson 2001}}

\citet{Nelson 2001} constructed the data cubes with different metallicities, and different assumptions about mass loss and angular moment loss for $Z=0.02$. They presented the conservative data cube at $Z=0.02$ with primary masses 0.8-50\,M$_{\sun}$ and showed the initial-parameter region of eight subtypes of case A evolution in this data cube. This work only focused on three subtypes (case AD, AR, and AS) for the low-mass binaries (primary masses between 0.45 and 4\,M$_{\sun}$) with the nonconservative assumption that both mass loss and angular momentum loss from the system. Comparison with their contours with conservative assumption, the main difference is that the ranges of initial orbital period of case AD, AR, and AS in this work are much larger than that in their contours, because the nonconservative assumption that mass and angular momentum loss can shrink the orbits of detached binaries with longer initial orbital period, making them evolve into contact. For the primary mass of 1\,M$_{\sun}$, as an example, detached binaries with $P_0<0.5$\,d can evolve into contact through case AD, AR, or AS with conservative assumption, while those with $P_0<5.5$\,d can evolve into contact through case AD, AR, or AS with the nonconservative assumption. In addition, the contours in this paper include the binary system with much less primary mass, and show that case AD may be the only subtype where detached binaries evolve into contact, which agrees with the suggestion given by \citet{Jiang 2012}.

\subsection{Comparison with the models of \citet{Yakut 2005}}

\citet{Yakut 2005} investigated the evolution of close binaries with conservative and non-conservative assumptions. They pointed out that the boundary between cases AR and AS depends not just on initial mass ratio and initial orbital period, but also on the initial mass of the primary. This agrees with the results in this paper, which show the initial primary mass has a strong influence on the initial parameter spaces of case AD, AR, and AS as shown in Fig 1. They also suggested that the nonconservative effects may reduce the range of case AS, but increase that of case AR. The initial-parameter spaces of cases AD, AR, and AS shown in this paper also agrees with their suggestion. In their non-conservative evolutionary models, both components were solved simultaneously, and this paper adopted the same way (the TWIN mode of the EV code) to calculate the evolution of binaries with the non-conservative assumption. Therefore, the code used in this paper can show the same evolutionary tracks of binary in the log $R$ vs. $M$ plane as that of the binary shown in their Fig. 5(a). The main difference between their paper and this paper is that this paper considered the binaries with the primaries less massive than 0.8\,M$_{\sun}$, and focused on comparing low-mass contact binaries produced by case AD, AR, and AS.

\subsection{Connections between W UMa binaries and case AD, AR, and AS}

Although case AD, AR, and AS can lead detached binaries to contact when both components are still main-sequence stars, it is still unclear what are their consequences, and whether they can produce W UMa binaries (observational counterparts of low-mass contact binaries). This work has investigated the possible connections between W UMa binaries and case AD, AR, and AS. The results in this paper show that low-mass contact binaries produced by case AR and AS have a reasonable agreement with observed W UMa binaries in the orbital period-effective temperature diagram and the age-mass(or orbital period) diagram, while the agreement is much poorer for case AD. This helps to support the suggestion that case AD will lead to a merger of two components on a quite short timescale through a common-envelope phase, while case AR and AS may lead to a relatively long-lived contact situation observed as W UMa binaries \citep{Nelson 2001, Eggleton 2010, Jiang 2012}. However, it should be noted that W UMa binaries are also expected to ultimately coalesce into single stars due to Darwin instability when the spin angular momentum of the system is more than a third of its orbital angular momentum \citep{Hut 1980, Rasio 1995, Li 2006, Jiang 2010}, or due to thermal instability when the primary attempts to cross the Hertzsprung gap \citep{Webbink 1976}. Therefore, case AD, AR, and AS may be progenitors of poorly understood stellar merger, blue stragglers, and FK Com stars.

\subsection{The origin of W UMa binaries}
The origin of W UMa binaries is still a matter of debate, and they can be originated during pre-main sequence phase \citep[e.g. fission process,][]{Roxburgh 1966, Bilir 2005}, or during main-sequence phase \citep[e.g. the detached-binary channel,][]{Vilhu 1982, Bradstreet 1994, Jiang 2014MN}. Their ages have been used to distinguish two channels. For example, W UMa found in old ($\sim10\,$Gyr) and intermediate-age ($\sim5$\,Gyr) clusters support the detached-binary channel, because their ages are much larger than their contact lifetime. However, it is difficult to distinguish the origin of young ($\sim0.5$\,Gyr) W UMa binaries based on their age, because young W UMa binaries can be produced by both channels \citep{Roxburgh 1966, Bilir 2005,Vilhu 1982, Jiang 2014MN}. The results in this paper show that case AR and AS can produce a short-period limit (corresponding to a lower mass limit) at almost any age, agreeing with the observation that the short-period limit of W UMa binaries exist in a different age (from $\sim0.3$\,Gyr to $\sim10$\,Gyr) cluster, even including young  ($\sim0.3-1$\,Gyr) clusters. Therefore, the short-period limit may be another important property to distinguish the origin of W UMa binaries, although it is still unclear whether fission process can also explain the existence of the short-period limit of W UMa binaries at young age.

\subsection{The progenitors of W UMa binaries}

The typical progenitors of the detached-binary channel are close binaries with initial orbital periods shorter than a certain critical value, e.g. the upper-period limit \citep{Vilhu 1982, Stepien 2011, Jiang 2014MN}. Identification of these progenitors would be used to study and support this channel. The results in this paper show that the mass ratio also has an important influence on identifying potential progenitors of W UMa binaries. Only the binaries with small mass ratio (i.e., equal masses) will experience case AR or AS evolution to form a W UMa binary, for example, V506 Oph \citep[$q=M_1/M_2\sim1.01, $][]{Torres 2019}. Even if some binaries with large mass ratio (i.e., unequal masses) have a period shorter than the upper-period limit, they will fail to produce a long-lived W UMa binaries due to case AD evolution, e.g. WY Cnc \citep[$P\sim0.83, q\sim2.61, $][]{Heckert 1998, Chen 2013}, CSTAR 036162 \citep[$P\sim0.87d, q\sim2.82, $][]{Liu 2018}. However, they may be the best candidates of the common-envelope merger, and can also be used to constrain adiabatic mass-loss models in binaries \citep{Ge 2010}. Therefore, the parameter spaces of case AD, AR, and AS can be expediently used to distinguish the progenitors of W UMa binaries or the common-envelope merger.

\subsection{The following evolution of low-mass contact binaries}

At present, this work studied only the formation of low-mass contact binaries from case AS and AR. Their following evolution and structure are still far from full understanding. The most difficult problem is the process of energy transfer between two components, which roughly equalizes the surface temperatures of two stars with different masses. This process has been supported by total observational luminosity of W UMa binaries equal to their total nuclear luminosity \citep{Mochnacki 1981, Webbink 2003, Jiang 2009}. Some physical model for this energy-transfer process has been developed \citep{Kahler 2002a, Kahler 2002b, Kahler 2004, Li 2004a, Li 2004b, Li 2005, Yakut 2005}. Differential rotation observed in the Sun has been suggested to be a new mechanism for the energy transfer between two components \citep{Yakut 2005, Eggleton 2010}. The energy-transfer rate is found to be related to the luminosity ratio or mass ratio \citep{Csizmadia 2004, Li 2008, Jiang 2009}, and has significant influence on the structure and evolution of contact binaries \citep{Li 2004a, Jiang 2009}. However, it is not considered that low-mass contact binaries may be originated differently by case AR or AS at present, and it is unclear whether low-mass contact binaries produced by case AR and AS have same energy-transfer mechanism, or same energy-transfer rate. More theoretical work is obviously needed to investigate the following evolution of low-mass contact formed from case AR and AS with considering energy-transfer process.

\section{Conclusions}
By performing detailed binary evolution calculations for low-mass close binaries with the nonconservative assumption,  three subtypes (AR, AS, and AD) of binary evolution channels were investigated, which may lead to contact while both components are still main-sequence stars. The results in this paper showed that the initial-parameter spaces of three subtypes strongly depend on the initial mass of the primary. Both the low-mass contact and their progenitors show a low-mass limit for case AR and AS, but not for case AD. The distributions of these low-mass contact binaries in period-effective temperature and the age-mass (or period) diagrams support the view that case AR and AS may form a W UMa binary, while case AD may lead to a binary-merger event through a rapid, unstable contact phase (or called " a common envelope phase"), instead of a stable contact binary. In future investigations, the structure and evolution of low-mass contact binaries produced by case AR and AS will be studied, which will help us to understand W UMa systems. The properties of case AD can also be used to find the observed candidate of the common-envelope merger.


\section*{Acknowledgements}
It is a pleasure to thank an anonymous referee for many valuable suggestions and comments, which improved the paper greatly. This work is supported by the Natural Science Foundation of China (Nos 11573061, 11733008, 11521303, 11773065 and 11661161016), by the Yunnan province (Nos 2015FB190, 2017HC018, 2013HA005). Guoshoujing Telescope (the Large Sky Area Multi-Object Fiber Spectroscopic Telescope, LAMOST) is a National Major Scientific Project built by the Chinese Academy of Sciences. Funding for the project has been provided by the National Development and Reform Commission. LAMOST is operated and managed by the National Astronomical Observatories, Chinese Academy of Sciences.







\bsp	
\label{lastpage}
\end{document}